\documentclass{aa}  

\usepackage{graphicx}
\usepackage{txfonts}
\begin{document}

   \title{Correlation between peak energy and Fourier power density spectrum slope in gamma--ray bursts}


   \author{S.~Dichiara\thanks{dichiara@fe.infn.it}
          \inst{1,2}
          \and
          C. Guidorzi\inst{1}
          \and
          L.~Amati\inst{3}
          \and
          F.~Frontera\inst{1,3}
          \and
          R.~Margutti\inst{4,5}
          }

   \institute{Department of Physics and Earth Sciences, University of Ferrara, via Saragat 1,
     I-44122, Ferrara, Italy\\
     \and
     ICRANet, P.zza della Repubblica 10, I-65122, Pescara, Italy\\
     \and
     INAF -- Istituto di Astrofisica Spaziale e Fisica Cosmica, Bologna, Via Gobetti 101, 
     I-40129 Bologna, Italy\\
     \and
     Harvard-Smithsonian Center for Astrophysics, 60 Garden St., Cambridge, MA 02138, USA\\
     \and
     New York University, Physics department, 4 Washington Place, New York, NY 10003, USA\\
%
%
   }


 
   \abstract
       {The origin of the gamma--ray burst (GRB) prompt emission still defies explanation, in spite of recent progress made, for example,
       on the occasional presence of a thermal component in the spectrum along with the ubiquitous non-thermal component that is modelled with a
       Band function.
       The combination of finite duration and aperiodic modulations make GRBs hard to characterise temporally.
       Although correlations between GRB luminosity and spectral hardness on one side and time variability on
       the other side have long been known, the loose and often arbitrary definition of the latter makes the
       interpretation uncertain.}
       {We characterise the temporal variability in an objective way and search for a connection with rest-frame
         spectral properties for a number of well-observed GRBs.}
       {We studied the individual power density spectra (PDS) of 123 long gamma-ray bursts with measured redshift,
         rest--frame peak energy $E_{\rm p,i}$ of the time--averaged $\nu$\,$F_\nu$ spectrum, and well--constrained
         PDS slope $\alpha$ detected with {\em Swift}, {\em Fermi} and past spacecraft.
         The PDS were modelled with a power law either with or without a break adopting a Bayesian Markov chain
         Monte Carlo technique.}
       {We find a highly significant $E_{\rm p,i}$--$\alpha$ anti-correlation. The null hypothesis probability 
           is $\sim 10^{-9}$.} 
       {In the framework of the internal shock synchrotron model, the $E_{\rm p,i}$--$\alpha$ anti-correlation can hardly be
         reconciled with the predicted $E_{\rm p,i}\propto\Gamma^{-2}$, unless either variable microphysical
         parameters of the shocks or continual electron acceleration are assumed.
         Alternatively, in the context of models based on magnetic reconnection, the PDS slope and $E_{\rm p,i}$
         are linked to the ejecta magnetisation at the dissipation site, so that more magnetised outflows would
         produce more variable GRB light curves at short timescales ($\lesssim1$~s), shallower PDS,
         and higher values of $E_{\rm p,i}$.}

   \maketitle
%

\section{Introduction}
\label{sec:intro}
The nature of the prompt gamma-ray emission remains one of the most elusive aspects of gamma-ray bursts (GRBs).
Most important questions concern the nature of the emitting ejecta and its
magnetisation degree, where and how energy dissipation takes place, and the
dissipation mechanism(s) (see \citealt{KumarZhang15rev,Zhang14a,Peer15rev,Zhang02} for reviews).

Typically, $10^{51}$--$10^{53}$ ~ergs are released as gamma-rays in a few seconds
within a region of about $10^7$~cm. The energy spectrum is highly non-thermal
and is usually described empirically by the Band function \citep{Band93},
a smoothly joined broken power-law whose $\nu$$F_\nu$ spectrum peaks at
$E_{\rm p}$ of a few hundred keV \citep{Preece00,Kaneko06,Guidorzi11c,Goldstein12,
Gruber14,Bosnjak14}, in some cases accompanied by a mostly subdominant thermal component
(e.g. \citealt{Guiriec11,Guiriec15}; see \citealt{Peer15rev} for a review).
An ultra-relativistic outflow (with Lorentz factor $\Gamma$ of a few $10^2$) is required to reduce the
pair production opacity and explain why a non-thermal
spectrum extending to the MeV and GeV ranges is observed
(e.g. \citealt{MeszarosGehrels12,Peer15rev}).
Generally, the different proposed dissipation processes can be classified based on
the distance from the source: i) the dissipation into gamma-rays takes place well above
the Thomson photosphere but still below the deceleration radius,
as in the case of the internal shock (IS) model \citep{Rees94,Narayan92};
ii) the dissipation occurs near the photosphere and
the emerging blackbody-like spectrum is distorted by additional heating and Compton scattering.
In either case the dissipation details depend on the ejecta magnetisation,
$\sigma=B^2/4\pi\Gamma\rho c^2$, defined as the ratio between the magnetic field
and matter energy densities, because it can affect the dynamic evolution of the outflow.
In models i) a fraction of energy is dissipated into radiation directly through IS,
or through magnetic reconnection in the case of a magnetised jet
\citep{Usov92,Thompson94,LyutikovBlandford03,McKinney12}.
In the case of classical IS the dissipated energy is kinetic of the baryon load; 
instead, for the magnetised outflows it is mostly Poynting flux (i.e. $\sigma\ge 1$ in the dissipation region).
In the ICMART model \citep{ICMART} magnetic reconnection is a consequence of the distortion of the magnetic
field lines entrained in the ejecta and triggered by IS (see \citealt{Kagan15} for a recent
review on relativistic magnetic reconnection). 
For models ii), energy dissipation takes place at or below the photosphere, either for baryonic-dominated outflows
\citep{Rees05,Peer06b,Thompson07,Derishev99,Rossi06,Beloborodov10,Titarchuk12},
or for magnetically dominated outflows \citep{Giannios08,Meszaros11,Giannios12}.
See \citet{Zhang14a}, \citet{Peer15rev} and \citet{Granot15rev} for recent reviews of the different models
of prompt emission.

In this context, timing analysis and its spectral characterisation can provide
insights into the size and distance of the dissipation region, and consequently
on the jet composition, the radiative processes and the geometry of the prompt emission,
since all these issues are intertwined (e.g. \citealt{BeniaminiPiran14}).
A classical way of characterising time variability of stochastic processes is
offered by Fourier analysis, and astrophysical time series are no exception
(e.g. see \citealt{Klis89,Vaughan13} for reviews).
The study of the continuum of the power density spectrum (PDS),
corresponding to the Fourier transform of the auto-correlation function
(ACF) of a time series, and the possible presence of periodic features in it
can constrain the spatial distribution of sources
contributing to the observed flux (e.g. \citealt{Titarchuk07}).
As found in several independent data sets \citep{Beloborodov00a,Ryde03,Guidorzi12,Dichiara13a,Vanputten14},
the  average PDS of long GRBs is described by a power law
extending over two frequency decades, from a few $10^{-2}$ to $1$-$2$~Hz.
The power-law index lies in the range $1.5$-$2$ with a small but significant
dependence on photon energy, with steeper slopes corresponding to softer
energy bands. Evidence for a break around $1$-$2$~Hz for the harder ($\ga 100$~keV) 
energy channels was also found \citep{Beloborodov00a, Dichiara13a}.

Whilst the  average PDS over a large number of GRB exhibits small fluctuations
and is easier to characterise in terms of a general stochastic process, it
provides no clues on the variety of properties of individual GRBs.
In this work we model the individual PDS and study the statistical
properties of an ensemble of GRBs that were detected by different past and present
spacecraft, and have measured redshift and a well-constrained
intrinsic (i.e. rest-frame) time-averaged spectral peak energy $E_{\rm p,i}$.
The difficulty of a proper statistical treatment of the PDS of an
individual, highly non-stationary, and short-lived stochastic process
such as that of a GRB time series, is properly overcome
with the aid of a Bayesian Markov chain Monte Carlo (MCMC) technique, which is
described in detail in a companion paper (\citealt{Guidorzi16}; hereafter, G16)
and is essentially the same as that outlined by \citet{Vaughan10}, except for a few
minor but important changes.
The same technique has recently been adopted for studying a selected sample
of bright short GRBs \citep{Dichiara13b} and outbursts
from soft-gamma ray repeaters \citep{Huppenkothen13}, for which Fourier
analysis faces the same formal problems of short-lived,
non-stationary time series.

The key point of studying individual vs. averaged PDS is that we can investigate
the possible connection between PDS and other key properties of the prompt
emission, such as $E_{\rm p,i}$, or the isotropic-equivalent radiated energy,
$E_{\rm iso}$, involved in the eponymous correlation \citep{Amati02}.
Establishing a connection between hydrodynamical and geometric quantities
and observables that characterise the spectral formation provides a powerful test
to constrain the prompt emission models.

The paper is organised as follows: the data selection and
analysis are described in Sect.~\ref{sec:data}. Sect.~\ref{sec:res}
reports the results, which are discussed in Sect.~\ref{sec:disc}.
The description of the technique adopted for the PDS modelling
is detailed in G16.
Uncertainties on the best-fitting parameters are 1~$\sigma$
confidence for one parameter of interest, unless stated otherwise.

\section{Data analysis}
\label{sec:data}

\subsection{Data selection}
\label{sec:data_sel}
From the data sets obtained with the main past and present experiments we selected the GRBs with measured
redshift and well-constrained $E_{\rm p,i}$.
Spectral parameters ($E_{\rm p,i}$ in particular) of past GRBs were taken from
the literature. For {\em Fermi}-catalogued GRBs we took the best-fit parameters
obtained by the team \citep{Goldstein12,Gruber14} when available.
For recent GRBs and/or {\em Fermi} GRBs with a shallow high-energy power-law
index $\beta_{\rm B}<-2$,\footnote{To ensure a finite maximum in the $\nu$$F_\nu$ spectrum.}
as modelled with the Band function \citep{Band93}, the parameters were
taken instead either from Konus-{\em WIND}, or, if not available, from the {\em Fermi}/GBM GCN.
We also calculated the isotropic-equivalent radiated energy $E_{\rm iso}$ in the rest-frame
1-$10^4$~keV band adopting $H_0=70$\,km\,s$^{-1}$\,Mpc$^{-1}$, $\Omega_\Lambda=0.70$,
$\Omega_{\rm M}=0.30$.

To ensure the best signal-to-noise (S/N) ratio
and lessen the effects of a limited energy passband, we took for
each instrument the time profile in the broadest energy range available.
For a subsample of GRBs detected with both {\em Fermi} and {\em Swift}
we discuss the impact of different energy bands and detectors. 
GRBs whose light curves were hampered by gaps in the time profiles were excluded.
We imposed a minimum threshold of 30 to the S/N as measured from
the net count fluence in the time interval selected for the PDS extraction
of each GRB. Finally, GRBs whose uncertainty on the PDS power-law index $\alpha$ was
$>$$0.5$ were rejected, except for five events for which the 90\%-confidence
lower limit to $\alpha$ is nonetheless very informative (Sect.~\ref{sec:pds_calc}).

\subsubsection{{\em Swift}-BAT data}
\label{sec:data_sel_swift}
From an initial sample of 961 GRBs detected by BAT
from January 2005 to May 2015 we selected those whose
time profiles are entirely covered in burst mode, that is,
those with the finest time resolution available. As a consequence,
the ground-discovered GRBs were excluded. We then extracted mask-weighted,
background-subtracted light curves with a uniform binning time of 4~ms
in the total passband 15-150~keV with the HEASOFT package (v6.13)
following the BAT team threads.\footnote{
http://swift.gsfc.nasa.gov/docs/swift/analysis/threads/bat\_threads.html}
Mask-weighted light curves were extracted using the ground-refined
coordinates provided by the BAT team for each burst through the
tool {\tt batbinevt}. We built the BAT detector quality map of each GRB
by processing the next earlier enable or disable map of the detectors.

We selected the long bursts by requiring $T_{90}>3$~s, where $T_{90}$
were taken from the second BAT catalogue \citep{Sakamoto11} or from the
corresponding BAT-refined circulars for the most recent GRBs not included in the catalogue.
We verified that no short GRB with extended emission \citep{Norris06,Sakamoto11}
with $T_{90}>3$~s slipped into the long-duration sample.
We ended up with 75~long {\em Swift}-BAT GRBs with measured redshift $z$ and $E_{\rm p,i}$.
Hereafter, we refer to these 75~long GRBs with measured quantities (PDS,
$z$, and $E_{\rm p,i}$) as the {\em Swift} BAT sample.

For complementary analysis we also considered the short GRB class. Only one short GRB, 051221A \citep{Burrows06,Soderberg06,Jin07}, matched all our criteria. We therefore considered it separately, for comparison reasons.

\subsubsection{{\em Fermi}-GBM data}
\label{sec:data_sel_fermi}
We selected a sub-sample of 102 GRBs detected by {\em Fermi}/GBM from July 2008 to May 1, 2015 with measured
redshift. For each GRB we took the time-tagged event (TTE) files of the two most illuminated NaI detectors
and extracted the corresponding light curves with 64~ms resolution in the 8-1000~keV band using the
{\tt gtbin} tool. We then summed them to increase the S/N.
We excluded all the GRBs that either had no TTE file or whose TTE data did not cover the whole event.
We excluded short-duration bursts by requiring $T_{90}>3$~s, and again we made sure not to include those
with extended emission.
Another selection was applied to remove all the light curves affected by spikes caused by
high-energy particles interacting with the spacecraft \citep{Meegan09}.

Data were processed by following the {\em Fermi} team
threads\footnote{\url{http://fermi.gsfc.nasa.gov/ssc/data/analysis/scitools/gbm\_grb\_analysis.html}}.
By virtue of its exceptional S/N, the light curve of 130427A was extracted with 10 ms resolution.
After further selection based on the PDS best-fit parameters (Sect.~\ref{sec:pds_calc}),
we ended up with 44 GRBs with redshift and $E_{p,i}$ that hereafter constitute the {\em Fermi} sample.

\subsubsection{Data from past experiments}
\label{sec:data_others}
Likewise, for the {\em BeppoSAX}/GRBM sample we started from the GRB catalogue
\citep{Frontera09} by selecting the GRBs covered with $7.8$ ms resolution,
available in the 40-700~keV energy band only for those that triggered the GRBM on-board logic.
In one case (010222) the time profile was not entirely covered with high resolution; by 
comparing the high-resolution PDS with the one extracted over the full profile at lower (1~s)
resolution, no remarkable difference was noted in the common frequency range, so we used
the high-resolution PDS because of the better sampled high-frequency tail.

For other experiments we took the time profiles in the following energy bands: 50-300~keV ({\em CGRO}-BATSE; 64~ms resolution),
\footnote{\url{ftp://cossc.gsfc.nasa.gov/compton/data/batse/ascii_data/64ms/}}
$>7$~keV (energy channels A+C+D of {\em HETE2}-FREGATE;
328~ms);\footnote{\url{http://space.mit.edu/HETE/Bursts/Data/}}
and 50-200~keV (Konus-{\em WIND}; 64~ms).\footnote{\url{http://gcn.gsfc.nasa.gov/konus_grbs.html}}
For all these instruments background subtraction was carried out through interpolation with up to
second-order polynomials.

We ended up with valuable data for 8, 7, 6, and 3 GRBs detected with {\em BeppoSAX}, BATSE,
{\em HETE2}, and Konus/{\em WIND}, respectively.

\subsection{PDS calculation}
\label{sec:pds_calc}
The time interval over which the PDS were calculated depended on the experiment.
For {\em Swift}-BAT we followed the same procedure as in \citealt{Guidorzi12}: we chose the $T_{7\sigma}$
interval, that is, the time interval whose boundaries correspond to the first
and last time bins whose rates exceed the background
level by $\ge7\,\sigma$. The PDS was then calculated on a $3\times\,T_{7\sigma}$ time interval
 that had the same central time.
As explained in \citealt{Guidorzi12}, this choice is a good trade-off between the need
for a full coverage of the GRB profile and that for an optimal S/N.
For most of the {\em Swift} GRBs $T_{7\sigma}$ is very similar to the more popular
$T_{90}$. Detailed examples of these individual PDS are shown in Fig.~1 of G16.
For the non-{\em Swift} data we calculated the PDS for each GRB in the
$T_{5\sigma}$ time interval \citep{Dichiara13a}.
The reason for the different choice lies in the more reliable {\em Swift}-BAT
background subtraction as a mask detector, whereas background subtraction
such as is obtained by interpolating counts of open-sky detectors over long intervals
 is likely to bias the low-frequency power of the PDS.
The {\em Fermi}-{\em Swift} common sample allows us to evaluate the effect 
of selecting different time intervals (Sect.~\ref{sec:res}).

All the PDS were calculated adopting the Leahy normalisation, in which the
constant power due to uncorrelated statistical noise has a value of
2 for pure Poissonian noise \citep{Leahy83}.

For each light curve we initially calculated the PDS by keeping the original
minimum binning time for each experiment.
After we made sure that no high-frequency ($f\ga 10$~Hz) periodic feature
stood out from the continuum, we decided to cut down the long computational
time demanded by Monte Carlo simulations by binning up the light curves of {\em Swift}-BAT
to $32$~ms, equivalent to a Nyquist frequency $f_{\rm Ny}=15.625$~Hz.
We did not adopt the potential alternative approach of binning up along
frequency, after we noted that the corresponding distribution of power at
high frequencies significantly deviated from the expected $\chi^2_{2M}$,
where $M$ is the re-binning factor \citep{Klis89}.
Unless stated otherwise, the PDS hereafter discussed were calculated in this
way. Neither white-noise subtraction nor frequency re-binning was applied
to the original PDS.
Table~\ref{tab:data} reports the time intervals used for the PDS calculation.

\subsection{PDS modelling}
\label{sec:pds_fit}
We modelled the observed PDS following the procedure presented in the method paper
(G16) based on a Bayesian Markov chain Monte Carlo technique, where two competing models are considered.
The simpler one is a mere power-law ({\sc pl}) plus the white-noise constant,
\begin{equation}
S_{\rm PL}(f) = N\,f^{-\alpha} + B\;,
\label{eq:mod_pl}
\end{equation}
where the following parameters were left free to vary:
the normalisation constant $N$, the power-law index $\alpha$ ($>0$), and the white-noise level $B$.
The PDS of several GRBs showed evidence of a break in the power law. For such cases, the
procedure considers a model of a power law with a break, below which the PDS is asymptotically constant,
which we here call bent power-law ({\sc bpl}) model,
\begin{equation}
S_{\rm BPL}(f) = N\,\Big[1 + \Big(\frac{f}{f_{\rm b}}\Big)^{\alpha}\Big]^{-1} + B\;,
\label{eq:mod_bpl}
\end{equation}
which reduces to the simple {\sc pl} model of Eq.~(\ref{eq:mod_pl}) in the limit $f\gg f_{\rm b}$
($\alpha>0$).
$f_{\rm b}$ is the break frequency, below which the power density flattens.
The preference for this model over the broken power-law model, such as that of Eq.(1) of \citealt{Guidorzi12}
used for the average PDS, is provided in paper~G16.
A justification for the specific choice of Eq.~(\ref{eq:mod_bpl}) is that it provides
a good description of the typical PDS of a fast rise exponential decay (FRED) pulse
(e.g. \citealt{Lazzati02}).

A likelihood ratio test is used to establish whether a {\sc bpl} provides a statistically
significant improvement in the fit of a given PDS, with a 1\% threshold on probability.
We refer to G16 for a detailed description and justification
of this method.

For GRBs with very good S/N and poor time resolution, such as {\em HETE2}-FREGATE whose Nyquist
frequency is a mere $1.5$~Hz, the white-noise level cannot be guessed from the GRB PDS.
We therefore preliminarily estimated it from the PDS extracted over a different
time interval adjacent to that of the GRB, including only background counts.
Under the reasonable assumption that the temporal properties of background counts did not
change, we constrained the white-noise level in the GRB PDS by means of a prior distribution for $B$.

\section{Results}
\label{sec:res}
We rejected all the GRBs with $\sigma(\alpha)>0.5$, except for five GRBs (970508, 980425, 990712,
021211, and 030528) for which the 90\%-confidence lower limit to $\alpha$ yielded a useful constraint. 
The selection on $\alpha$ was decided considering its observed range (1.3<$\alpha$<3.9).
This threshold allowed us to constrain the PDS slopes in a meaningful way.
We ended up with 123 different GRBs with all measured and constrained $E_{\rm p,i}$, $E_{\rm iso}$, 
and $\alpha$ (and $\log{f_b}$ for the GRBs with a break in the PDS), plus just one short from the 
{\em Swift} sample (Sect.~\ref{sec:data_sel_swift}).
Twenty GRBs detected with both {\em Fermi} and {\em Swift} passed
the PDS selection with both data sets. We used this sample to explore the effects of different energy bands
and different time intervals.
Best-fitting parameters are reported in Table~\ref{tab:data}\footnote{Tables 1 is only available in electronic form at the CDS via anonymous ftp to cdsarc.u-strasbg.fr (130.79.128.5) or via http://cdsweb.u-strasbg.fr/cgi-bin/qcat?J/A+A/}. Some GRBs that were detected by both
spacecraft are not classified as common, since only the data of one of the detectors matched our selection
criteria. In these cases we reported the experiment whose data gave a useful PDS.
%
\begin{table*}
\centering
\caption{Table of the sample of 123 GRBs. The PDS is calculated in the time interval reported.
This table is available in its entirety in a machine-readable form in the online journal. A portion is shown here for guidance.}
\label{tab:data}
\begin{tabular}{lllccrrccrr}
\hline
GRB & E$^{\mathrm{(a)}}$ & $z$ & $\log{E_{\rm iso}}$ & $\log{E_{\rm p,i}}$ & $t_{\rm start}^{\mathrm{(b)}}$ & $t_{\rm stop}^{\mathrm{(b)}}$ & $\alpha$ & $\log{f_b}^{\mathrm{(c)}}$ & Ref$^{\mathrm{(d)}}$ &Ref$^{\mathrm{(e)}}$\\
    &    &       & ($10^{52}$~erg)  & (keV) & (s) & (s) & & (Hz) &$z$ & $E_{\rm p,i}$ \\
\hline
970228  & BS    & $0.695$ & $ 0.216\pm 0.033$ & $ 2.265\pm 0.148$ & $-0.1$ & $70.9$ & $ 3.51\pm 0.35$ & $-0.52\pm 0.08$ &  (1) &  (1)\\
970508  & B     & $0.835$ & $-0.209\pm 0.093$ & $ 2.141\pm 0.133$ & $-0.6$ & $10.0$ & $ 4.95\pm 1.60$ & NA &  (1) &  (1)\\
970828  & B     & $0.958$ & $ 1.480\pm 0.051$ & $ 2.759\pm 0.088$ & $0.0$ & $120.8$ & $ 2.12\pm 0.06$ & NA &  (1) &  (1)\\
971214  & BS    & $3.42$ & $ 1.340\pm 0.055$ & $ 2.827\pm 0.085$ & $-2.0$ & $30.9$ & $ 1.57\pm 0.12$ & NA &  (1) &  (1)\\
980425  & B     & $0.0085$ & $-4.000\pm 0.080$ & $ 1.706\pm 0.175$ & $-1.8$ & $225.2$ & $ 4.93\pm 1.82$ & NA &  (1) &  (1)\\
980703  & B     & $0.966$ & $ 0.868\pm 0.042$ & $ 2.698\pm 0.056$ & $-26.6$ & $66.8$ & $ 3.25\pm 0.33$ & NA &  (1) &  (1)\\
990123  & BS    & $1.6$ & $ 2.376\pm 0.071$ & $ 3.220\pm 0.120$ & $-7.6$ & $75.1$ & $ 2.29\pm 0.08$ & NA &  (1) &  (1)\\
990506  & B     & $1.3$ & $ 1.990\pm 0.044$ & $ 2.819\pm 0.102$ & $-2.0$ & $205.5$ & $ 2.53\pm 0.05$ & $-0.94\pm 0.06$ &  (1) &  (1)\\
990510  & B     & $1.619$ & $ 1.253\pm 0.066$ & $ 2.624\pm 0.043$ & $-0.6$ & $108.1$ & $ 2.13\pm 0.05$ & NA &  (1) &  (1)\\
990705  & BS    & $0.842$ & $ 1.253\pm 0.062$ & $ 2.641\pm 0.136$ & $-0.2$ & $41.2$ & $ 2.36\pm 0.14$ & $-0.36\pm 0.09$ &  (1) &  (1)\\
990712  & BS    & $0.434$ & $-0.172\pm 0.085$ & $ 1.963\pm 0.071$ & $1.0$ & $18.6$ & $ 3.30\pm 0.58$ & NA &  (1) &  (1)\\
991208  & K     & $0.706$ & $ 1.360\pm 0.035$ & $ 2.493\pm 0.043$ & $-0.1$ & $72.6$ & $ 2.45\pm 0.09$ & NA &  (1) &  (1)\\
991216  & BS    & $1.02$ & $ 1.842\pm 0.045$ & $ 2.802\pm 0.091$ & $0.5$ & $25.4$ & $ 3.05\pm 0.11$ & $-0.10\pm 0.06$ &  (1) &  (1)\\
000131  & B     & $4.5$ & $ 2.257\pm 0.077$ & $ 2.952\pm 0.195$ & $0.1$ & $120.5$ & $ 2.33\pm 0.07$ & NA &  (1) &  (1)\\
000418  & K     & $1.12$ & $ 0.970\pm 0.083$ & $ 2.452\pm 0.032$ & $-0.4$ & $31.4$ & $ 1.82\pm 0.17$ & NA &  (1) &  (1)\\
000911  & K     & $1.06$ & $ 1.835\pm 0.090$ & $ 3.260\pm 0.088$ & $-0.3$ & $26.0$ & $ 1.42\pm 0.12$ & NA &  (1) &  (1)\\
010222  & BS    & $1.48$ & $ 1.926\pm 0.046$ & $ 2.884\pm 0.017$ & $-67.0$ & $151.0$ & $ 2.43\pm 0.07$ & $-0.87\pm 0.08$ &  (1) &  (1)\\
011121  & BS    & $0.36$ & $ 0.884\pm 0.122$ & $ 3.010\pm 0.115$ & $-11.6$ & $259.3$ & $ 1.62\pm 0.04$ & NA &  (1) &  (1)\\
020813  & H     & $1.25$ & $ 1.821\pm 0.111$ & $ 2.756\pm 0.114$ & $0.2$ & $127.0$ & $ 2.25\pm 0.09$ & NA &  (1) &  (1)\\
021211  & H     & $1.01$ & $ 0.063\pm 0.050$ & $ 2.064\pm 0.189$ & $0.2$ & $7.4$ & $ 3.68\pm 0.56$ & NA &  (1) &  (1)\\
030226  & H     & $1.98$ & $ 1.102\pm 0.047$ & $ 2.449\pm 0.101$ & $-23.2$ & $73.9$ & $ 1.89\pm 0.38$ & NA &  (1) &  (1)\\
030328  & H     & $1.52$ & $ 1.588\pm 0.041$ & $ 2.510\pm 0.074$ & $-55.6$ & $164.6$ & $ 2.41\pm 0.20$ & NA &  (1) &  (1)\\
030329  & H     & $0.17$ & $ 0.163\pm 0.077$ & $ 1.988\pm 0.102$ & $-11.6$ & $142.9$ & $ 2.66\pm 0.07$ & NA &  (1) &  (1)\\
030528  & H     & $0.782$ & $ 0.343\pm 0.032$ & $ 1.751\pm 0.068$ & $0.3$ & $90.1$ & $ 4.88\pm 1.77$ & NA &  (1) &  (1)\\
050401  & S     & $2.90$ & $ 1.566\pm 0.086$ & $ 2.657\pm 0.104$ & $-42.6$ & $64.4$ & $ 2.31\pm 0.23$ & $-1.33\pm 0.28$ &  (1) &  (1)\\
050525A & S     & $0.606$ & $ 0.402\pm 0.076$ & $ 2.117\pm 0.013$ & $-12.6$ & $25.6$ & $ 3.13\pm 0.14$ & $-0.82\pm 0.11$ &  (1) &  (1)\\
\ldots   & \ldots & \ldots & \ldots   & \ldots & \ldots &\ldots   & \ldots & \ldots &\ldots   & \ldots \\
080916A & FS $^{\mathrm{(f)}}$ & $0.689$ & $ 0.052\pm 0.057$ & $ 2.328\pm 0.092$ & $-0.5$ & $52.5$ & $ 2.86\pm 0.43$ & NA &  (6) &  (3)\\
 & & & & & $-91.4$ & $177.0$ & $ 3.42\pm 0.40$ & $-1.53\pm 0.12$ &  & \\
080916C & F  & $4.35$ & $ 2.672\pm 0.006$ & $ 3.548\pm 0.030$ & $-3.8$ & $86.3$ & $ 1.96\pm 0.13$ & NA &  (7) &  (3)\\
\ldots   & \ldots & \ldots & \ldots   & \ldots & \ldots &\ldots   & \ldots & \ldots &\ldots   & \ldots \\
150323A & S  & $0.593$ & $0.114\pm 0.038$  & $2.182\pm 0.066$  & $-158.2$  & $324.7$ & $2.67\pm 0.17$ & $-1.78\pm 0.12$ & (69) & (33)\\
150403A & FS $^{\mathrm{(f)}}$ & $2.06$ & $ 2.000\pm 0.044$ & $ 3.053\pm 0.062$ & $-0.8$ & $42.1$ & $ 2.49\pm 0.19$ & NA & (70) & (34)\\
 & & & & & $-194.7$ & $217.5$ & $ 3.23\pm 0.20$ & $-1.45\pm 0.08$ &  & \\
\hline
\end{tabular}
\begin{list}{}{}
\item[$^{\mathrm{(a)}}$]{Experiments: BS={\em BeppoSAX}, B=BATSE, K=KONUS, H={\em HETE2}, S={\em Swift}, F={\em Fermi},
   FS={\em Fermi} and {\em Swift}.}
\item[$^{\mathrm{(b)}}$]{Times are referred to the trigger time.}
\item[$^{\mathrm{(c)}}$]{It is available for the GRBs whose PDS are best fit with a {\sc bpl} rather than a {\sc pl} model.}
\item[$^{\mathrm{(d)}}$]{(1) \citet{Amati08} and references therein. The full list is available on line. }
\item[$^{\mathrm{(e)}}$]{(1) \citet{Amati08} and references therein. The full list is available on line.}
\item[$^{\mathrm{(f)}}$]{(1) GRBs for which both {\em Fermi} and {\em Swift} data gave acceptable results are reported in two consecutive
  lines, referring to {\em Fermi} and {\em Swift}, respectively.}
\end{list}
\end{table*}
%
%
\begin{figure*}
\sidecaption
\includegraphics[width=12cm]{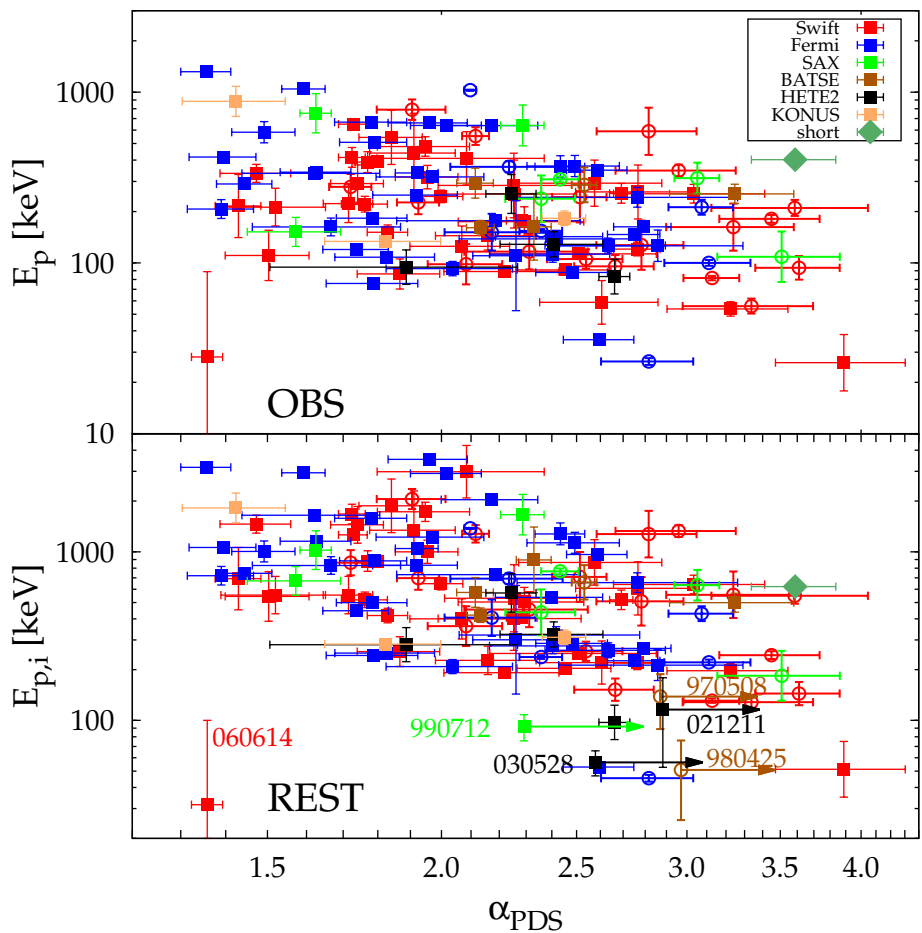}
\caption{Peak energy vs. PDS slope in the observer ({\em top}) and in the source
rest-frame ({\em bottom}). Different colours correspond to different experiments.
Filled squares (empty circles) correspond to power-law (broken power-law) models.
Lower limits on $\alpha$ are given at 90\% confidence.
The only short available is also shown for comparison (diamond).}
\label{fig:corr}
\end{figure*}
%

Out of the observables and best-fitting parameters available we found a highly significant correlation between
$E_{\rm p,i}$ and $\alpha$, as displayed in the bottom panel of Fig.~\ref{fig:corr}.
Unsurprisingly, $\alpha$ also correlates with $E_{\rm iso}$; however, the scatter is significantly larger, therefore hereafter we focus on $E_{\rm p,i}$--$\alpha$.
For comparison, we also studied the observed peak energy, $E_{\rm p}=E_{\rm p,i}/(1+z)$, vs. $\alpha$, shown in
the top panel of Fig.~\ref{fig:corr}.
For the 20 common GRBs we systematically chose the {\em Fermi} values for the broader energy passband. 
On average, the GRBs with higher $E_{\rm p,i}$ exhibit lower PDS indices.
The p-values associated with Pearson, Spearman, and Kendall
coefficients are $9\times10^{-8}$, $1\times10^{-8}$, and
$2\times10^{-8}$, respectively. These values do not account for the measurement
uncertainties. We conservatively
estimated their effect through MC simulations: we independently scattered
each point assuming a log-normal distribution along $E_{\rm p,i}$ and
using the marginal posterior distribution obtained for $\alpha$ for each GRB
(see G16). We generated 1000 synthetic sets
of 123 GRBs each and calculated the corresponding correlation coefficients.
The 90\% percentile values of the corresponding p-value distribution
are reported in Table~\ref{tab:corr} both for the rest- and for the observer
frame. 
The PDS index significantly correlates with both $E_{\rm p,i}$ and $E_{\rm p}$.
However, the significance of the correlation moves from
$10^{-5}$ to $10^{-7}$-$10^{-8}$ passing from observed to intrinsic plane,
improving by almost three orders of magnitudes.
We applied the same analysis to the subsample of {\sc pl} GRBs alone to investigate whether
the correlation still holds. We found that it does, with a p-value of the order of $10^{-6}$ in
the source frame. Like for the overall sample, moving from the source to the observer frame
the correlation becomes less significant by almost two orders of magnitude in this case
(see Table~\ref{tab:corr}).

Gamma-Ray burst 060614 is worth mentioning. This is a clear outlier
in a region of its own. This GRB is known to be peculiar for other
{\em observationally independent} reasons: although it is classified as a long GRB, its
nature yet remains ambiguous, since it shares some properties with
short bursts, such as the absence of any {associated supernova despite the small distance}, the
temporal lag of the initial spike \citep{Gehrels06,DellaValle06,Fynbo06}, and
the possible evidence for a macronova in the afterglow \citep{Yang15,Kisaka15,Jin15}.
Its prompt light curve consists of an initial hard spike followed by a soft
variable tail, and it was shown to be consistent with the $E_{\rm p,i}$-$E_{\rm iso}$
relation satisfied by most long GRBs only when the whole event is considered \citep{Amati07}.
By contrast, when only the negligible-lag spike is considered, its spectral and
temporal properties are more reminiscent of short GRBs \citep{Gehrels06,Amati07}.
If 060614 is excluded, the significance of the $\alpha$-$E_{\rm p,i}$ correlation
improves by a factor of a few (Table~\ref{tab:corr}).

We also studied the covariance of the PDS slope with the spectral indices associated
with the peak energy, derived from either a Band function or a cutoff power-law, and
found no departure from statistical independence.
%
\begin{table}  
  \tabcolsep 4pt         
  \begin{center}         
    \caption{$E_{\rm p,i}$-($E_{\rm p}$-) $\alpha$ correlation significance. Uncertainties on individual
      quantities have been accounted for.}
    \label{tab:corr}
    \begin{tabular}{llccc}
      \hline 
      \hline 
      \noalign{\smallskip} 
      Sample  &  $P$(Pearson) & $P$(Spearman)  & $P$(Kendall)\\
      \noalign{\smallskip}
      \hline
      $\alpha$ vs. $E_{\rm p}$ & $1\times10^{-5}$ & $2\times10^{-5}$ & $2\times10^{-5}$\\ 
      $\alpha$ vs. $E_{\rm p,i}$ & $2\times10^{-7}$ & $6\times10^{-8}$ & $8\times10^{-8}$\\ 
      $\alpha$ vs. $E_{\rm p,i}$\tablefootmark{(a)} & $2\times10^{-9}$ & $4\times10^{-9}$ & $1\times10^{-8}$\\ 
      $\alpha$ vs. $E_{\rm p}$\tablefootmark{(b)} & $8\times10^{-5}$ & $2\times10^{-4}$ & $1\times10^{-4}$\\ 
      $\alpha$ vs. $E_{\rm p,i}$\tablefootmark{(b)} & $6\times10^{-6}$ & $6\times10^{-6}$ & $6\times10^{-6}$\\ 
      \hline
      $\alpha$ vs. $E_{\rm p,i}$\tablefootmark{(c)} & $7\times10^{-2}$ & $9\times10^{-2}$ & $8\times10^{-2}$\\ 
      $\alpha$ vs. $E_{\rm p,i}$\tablefootmark{(d)} & $8\times10^{-4}$ & $4\times10^{-4}$ & $1\times10^{-3}$\\
      $\alpha$ vs. $E_{\rm p,i}$\tablefootmark{(e)} & $1\times10^{-5}$ & $1\times10^{-5}$ & $3\times10^{-6}$\\
      \noalign{\smallskip}
      \hline
    \end{tabular}
  \end{center}
  \tablefoot{
    \tablefoottext{a}{060614 excluded.}
    \tablefoottext{b}{Only GRBs whose PDS are best fit with {\sc pl}.}
    \tablefoottext{c}{Low S/N subsample (39 GRBs; SSN1)}
    \tablefoottext{c}{Mid S/N subsample (39 GRBs; SSN2)}
    \tablefoottext{c}{High S/N subsample (39 GRBs; SSN3)}
  } 
\end{table} 
%

\subsection{Dependence of PDS on energy band and on time interval}
\label{sec:shared}
\begin{figure}
\centering
\includegraphics[width=7cm]{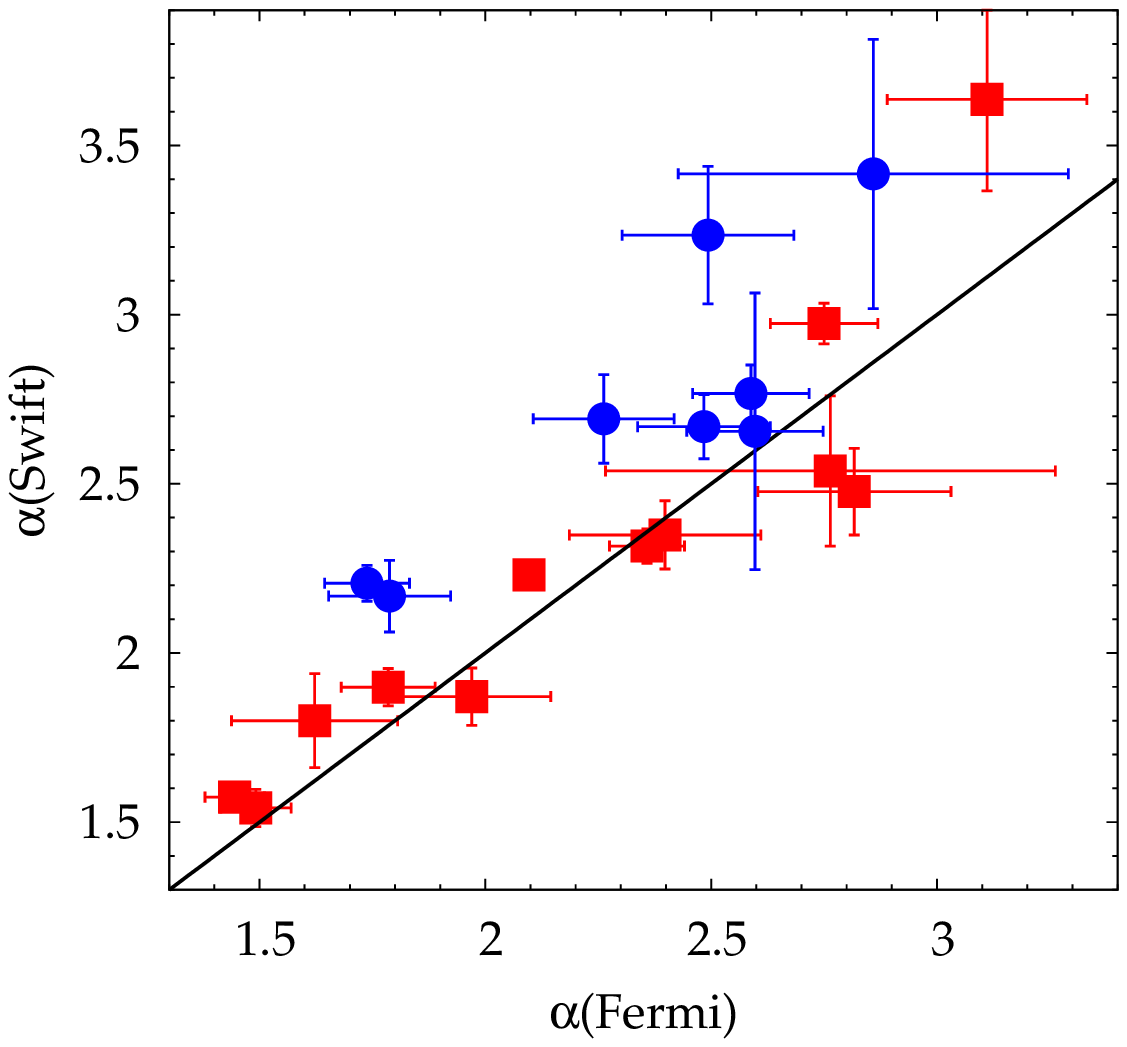}
\caption{PDS slope as measured with {\em Swift}-BAT and
and with {\em Fermi}-GBM data for the common GRBs.
Equality is shown for comparison (solid line).
The GRBs whose best-fit models are equal (different) for
both data sets are shown with squares (circles).}
\label{fig:shared}
\end{figure}

Fig.~\ref{fig:shared} compares $\alpha$ as measured with {\em Fermi}
and with {\em Swift} for the common sample of 20 GRBs. 

%
%
The two main differences between {\em Fermi}/GBM and {\em Swift}/BAT are the broader and harder energy band for
the former, along with a shorter time interval for the PDS calculation (Sects.~\ref{sec:data_sel} and~\ref{sec:pds_calc}).
The role of the energy passband on determining the slope of GRB PDS is already known
on the average PDS, with harder channels having lower $\alpha$'s because of the
more pronounced narrowness of their time profiles \citep{Dichiara13a}.
We therefore studied whether there is any link between the diversity of the two
measures and the GRB hardness, as measured with $E_{\rm p,i}$. We found no clear
evidence for it. Rather than the GRB hardness itself, it is the combination of a longer
time interval and a softer energy band of {\em Swift}-BAT that in most cases
turns into the identification of a low-frequency break in the PDS.
To clearly show this dependence, Fig.~\ref{fig:shared} displays the pairs of estimates with
two different symbols, depending on whether the two best-fit models coincide. In all
cases with different models, {\em Swift} data are best fit with a {\sc bpl} while
{\em Fermi} with a {\sc pl}. These events, shown with circles in Fig.~\ref{fig:shared},
deviate on average from equality more than the other GRBs, with higher values for
{\em Swift}, as expected from the properties of average PDS \citep{Dichiara13a}.

Yet, while the previously known average behaviour of PDS with energy is confirmed
in the bulk properties of our common sample, we cannot infer a universal behaviour for
all individual cases: for most common GRBs in Fig.~\ref{fig:shared} {\em Fermi} and
{\em Swift} PDS slopes are equal within uncertainties. This validates our choice of merging results
obtained with different experiments into a unique set (Fig.~\ref{fig:corr}).
In conclusion, merging results obtained in different energy bands does not wash out the
$E_{\rm p,i}-\alpha$ correlation, but probably contributes to the observed scatter. 

\subsection{Selection effects}
\label{sec:sel_eff}
\begin{figure}
\centering
\includegraphics[width=8.5cm]{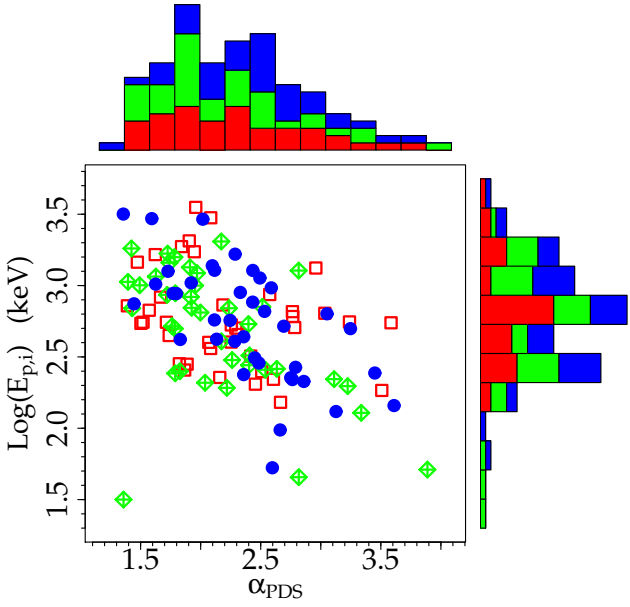}
\caption{$E_{\rm p,i}-\alpha$ for three different classes of the light curve S/N.
Blue circles, green diamonds, and red squares have high, middle, and low S/N, respectively.
The corresponding histograms are shown along both axes.
Error bars are not shown for the sake of clarity.
}
\label{fig:snr_subsamples}
\end{figure}

Since the correlation is highly scattered, we investigated the 
impact of selection effects that are due to S/N on the observed distribution of data on the 
$E_{\rm p,i}-\alpha$ plane.
Ideally, the S/N should be independent of both $E_{\rm p,i}$ and $\alpha$.
We therefore split the full sample into three different classes of S/N  
of 39 GRBs each with low (S/N < 52; hereafter SSN1), medium (52 < S/N < 121; SSN2), and high 
(S/N > 121; SSN3) values, and compared the distributions for the two variables in each subsample.
We left out the five cases with lower limits derived for $\alpha$ plus one GRB from 
SSN3 to ensure that all samples had the same statistical accuracy from the same number of GRBs.
We used the non-parametric Epps–Singleton (ES) test \citep{EppsSingleton86} to compute the 
probability that the $E_{\rm p,i}$ and $\alpha$ distributions of each sub-sample were drawn 
from a common one.
We opted for ES instead of the more popular KS, because the latter is less sensitive than the former,
especially for small samples.
For $E_{\rm p,i}$ we found probabilities of 2\%, 30\%, and 43\% for the comparisons SSN1-SSN2, SSN1-SSN3,
and SSN2-SSN3 sub-sets, respectively. Thus we can state that the three $E_{\rm p,i}$ 
distributions are consistent with being drawn from a common population. Similar conclusions are 
drawn about $\alpha$, with 82\%, 38\%, and 12\% analogues probabilities, respectively.
Finally, we computed the $E_{\rm p,i}-\alpha$ correlation for the three samples (see Table~\ref{tab:corr}) 
and found that the higher the S/N, the more significant the correlation. In other words, the better the 
signal, the less scattered the correlation, which means that the observed scatter is not entirely intrinsic to the sources,
but is also due to the uncertainties affecting the individual measurements. This is clearly illustrated 
in Fig.~\ref{fig:snr_subsamples} where we show the $E_{\rm p,i}-\alpha$ distribution for the three S/N sub-samples
along with their corresponding marginal distributions.
We therefore conclude that the correlation is not spurious and cannot artificially be caused by S/N effects.

%
\section{Discussion}
\label{sec:disc}
The choice of focusing on GRBs with well-measured $E_{\rm p,i}$ is driven by the
key role of this observable in understanding the mechanism of the prompt emission.
This is supported by a number of correlations in which $E_{\rm p,i}$ is involved:
its correlation with $E_{\rm iso}$ \citep{Amati02}, its time-resolved analogue
\citep{Golenetskii83,Yonetoku04,Ghirlanda11,Lu12b,Frontera12}, with the collimation-corrected
energy $E_{\gamma}$ \citep{Ghirlanda04} for long GRBs, and both $E_{\rm iso}$ and
the energy released in the X-ray band, $E_{\rm X,iso}$, in the
three-parameter correlation, which holds for both long and short GRBs
\citep{Bernardini12,Margutti13}. Thus, investigating the connection between $E_{\rm p,i}$ and 
temporal properties can provide clues on the physics of the prompt emission.

\begin{figure}
\centering
\includegraphics[width=8.5cm]{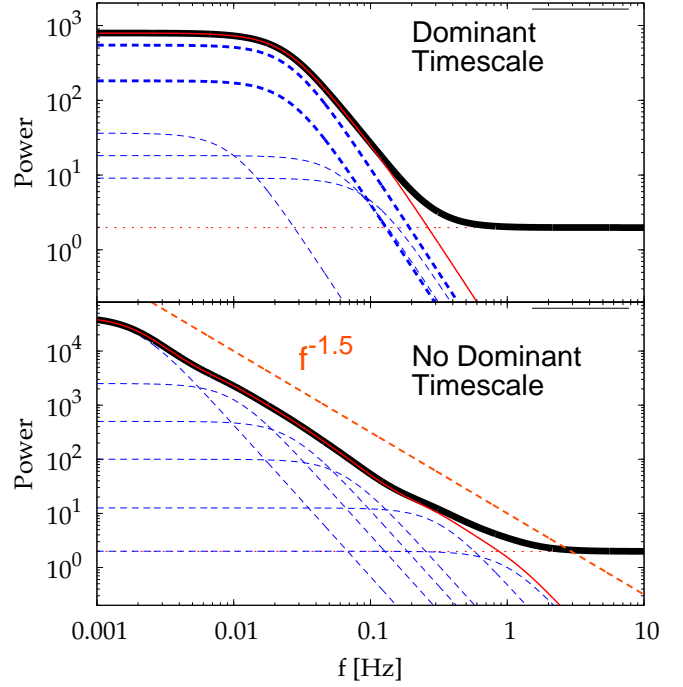}
\caption{{\em Top panel}: sketch of a {\sc bpl} PDS (thick solid line) as the
result of the superposition of PDS of different pulses (thin dashed lines).
The overall variance is dominated by pulses with similar timescales (thick dashed
lines), whose frequency break there corresponds to the dominant time.
The white-noise level is also shown (dotted line).
{\em Bottom panel}: the {\sc pl} PDS is the result of
the superposition of different pulses with different timescales, so that
no break stands out in the total PDS, which looks like a power-law with
a shallow index ($\alpha=1.5$ in this example, thick dashed line).}
\label{fig:sketch}
\end{figure}
%

In Fig.~\ref{fig:sketch} we illustrate our interpretation of the difference
between the two groups of PDS best fit with either {\sc bpl} or {\sc pl} and the meaning of
dominant timescale, wherever there exists one. Generally, a light curve is the result
of superposing a number of pulses with different timescales. Whenever the total
variance is dominated by some specific timescale, this stands out and determines the
break in the PDS, which is best fit with {\sc bpl} top panel of Fig.~\ref{fig:sketch}).
By contrast, when several different timescales have similar weights in the
total variance, the resulting PDS exhibits no clear break, and appears to be
remarkably shallower ($\alpha_{\rm pl}\la 2$) than that of individual pulses
($\alpha_{\rm bpl}>2$).
Most of the break frequency values correspond to timescales that are several second long.
This suggests that most GRBs with a break in the PDS lack power at sub-second scales,
which is not the case for the light curves of the {\sc pl} group.
Therefore, we interpret the two groups based on the relative strength of the short-timescale ($<1$~s; fast) component with respect to the long-timescale ($>1$~s; slow):
GRBs of the {\sc bpl} group exhibit weak or absent sub-second variability, whereas this is stronger in the {\sc pl} group.
Past works have reported evidence for two distinct (fast and slow) components
in GRB light curves, with the fast component being stronger in the harder energy bands
\citep{Shen03,Vetere06,Gao12}.
The correlation we found between PDS slope and $E_{\rm p,i}$ shows that GRBs with pronounced
sub-second variability in addition to the slow component have, on average, higher peak energies. 

The implications of our result can be discussed in the framework of some of the
main prompt emission models.
Considering the IS model, at given shell widths d and separations D, the higher the typical Lorentz factor, the correspondingly
larger the IS radius $r_{\rm is}\sim \Gamma^2 D$, while the observed timescales still reflect the
intrinsic inner engine variability.
The easiest interpretation of our result is to invoke the bulk Lorentz factor as the key observable
that explains the correlation between PDS and $E_{\rm p,i}$.
On the temporal side, we need to determine how $\Gamma$ may be related to the fast component in the observed light curves.
In early attempts to explain the luminosity-variability correlation
\citep{Fenimore00,Reichart01,Guidorzi05b,Rizzuto07}, it was shown that the $e^{\pm}$ pair photosphere
created by the IS synchrotron photons contributes to suppress the short-timescale variability
in GRBs with lower Lorentz factors \citep{Kobayashi02,Meszaros02}. The radius of this photosphere is $R_{\pm} \propto \sqrt{\Gamma_{\rm max}^{-1/2}\Gamma_{\rm min}^{-1/2}}$ ($\Gamma_{\rm min}$ and $\Gamma_{\rm max}$ are the Lorentz factors of the slowest and of the fastest shell). 
As a result, for $R_\pm > \Gamma_{\rm min}^2 D_{\rm max}$ ($D_{\rm min}$ and $D_{\rm max}$ are the smallest and
the largest separation between shells) the first collisions for most shells take place below the pair photosphere.
This turns into a wind of shells with ordered values of $\Gamma$ that come out of the photosphere and
can no longer effectively collide and produce bright and short pulses. On the other side,
for $R_\pm< \Gamma_{\rm min}^2 D_{\rm min}$ most collisions occur above the photosphere, thus preserving the
possibility of bright and short pulses in the observed light curve.
The natural radiation mechanism in the optically thin region is synchrotron; in this case,
$E_{\rm p,i}$ is the synchrotron peak energy and scales as $E_{\rm p,i}\propto\gamma_e^2\Gamma\,B\propto\,L^{1/2}\,r_{\rm is}^{-1}$,
where $\gamma_e$ and $B$ are the electron Lorentz factor and the magnetic field in the fluid comoving frame, and $L$ the
luminosity.
Since $r_{\rm is}\propto\Gamma^2$, the IS synchrotron model predicts $E_{\rm p,i}\propto\Gamma^{-2}$ \citep{Zhang02b,RamirezRuiz02},
which clashes with our result.
However, if $\Gamma$ correlates with the jet-viewing angle, so that $L\propto\Gamma_{\rm max}$, it is $E_{\rm p,i}\propto\Gamma^{-3/2}$
as long as the fractions of internal energy taken by the magnetic field, $\epsilon_B$, and by the accelerated electrons, $\epsilon_e$,
and the relative Lorentz factor of the colliding shells are constant. By letting them vary, it is possible to produce
$E_{\rm p,i}\propto\Gamma^{1/2}$ \citep{RamirezRuiz02}.
Therefore, while our result cannot rule out the IS synchrotron model, its interpretation dictates specific constraints
either in terms of continual acceleration of electrons in the shocked region, or on the dependence of $\epsilon_B$ and
$\epsilon_e$ on $\Gamma$.

In the photospheric models the dissipation takes place below the
photospheric radius at optical depths of a few to a few tens. Without going into the
details of the various dissipation processes, we can identify two different origins for
the prompt emission variability: (i) imprinted by the inner engine and (ii) resulting from
the interaction of the jet with the stellar envelope and the cocoon
\citep{MacFadyen99,Zhang03b,Lazzati05b,Morsony10}.
The basic idea to explain our result still relies on the possibility that the inner engine can
produce variable winds on sub-second timescales.
If the spread in the observed $E_{\rm p,i}$ distribution is mainly driven by a geometric
effect caused by different viewing angles $\theta_{\rm v}$ and, therefore, correspondingly different photospheric
radii, then lower peak energies, on average, correspond to larger $\theta_{\rm v}$ \citep{Lazzati11b,Lazzati13a}.
Our result suggests that the more off axis the observer, the weaker the expected high-frequency ($>1$\,Hz)
temporal power in the light curves. From detailed simulations that follow the jet
propagation from inside the star all the way out to the photospheric radius, while the peak luminosity
decreases at increasing $\theta_{\rm v}$, the high-frequency continuum power is generally described by a $\nu^{-2}$
behaviour and seemingly insensitive to $\theta_{\rm v}$ \citep{LopezCamara14}.

When at large distances (e.g. $\sim10^{15}$~cm in the ICMART model) the magnetisation is still
high ($\sigma\gtrsim 1$), magnetic reconnection is more efficient than shocks in dissipating
energy and accelerating non-thermal particles, as shown by recent particle-in-cell simulations
\citep{Sironi15}.

In the context of magnetic reconnection models for the GRB prompt emission, the essential idea that explains the
observed variability is the relativistic motion ($\gamma'\lesssim 10$) of emitting clumps (or eddies, or fundamental emitters)
within the bulk comoving frame.
Since magnetic reconnection can be the source of turbulence (e.g. \citealt{Lazarian15}),
the emitting eddies have turbulent motions within the comoving frame of the outflow, as prescribed in the relativistic
turbulence model \citep{Lazar09,Narayan09,Kumar09},
At variance with the previous models, fast variability originates in the emission region \citep{LyutikovBlandford03,Lyutikov06}.
The relative strength of the fast over the slow component reflects the filling factor
of the eddies in the jet. Within this picture, our result suggests that GRBs with a large (low) filling factor,
correspond to the high (low) $E_{\rm p,i}$ values. 
A possible problem with relativistic turbulence is that a large part of energy has to be continuously
maintained in turbulent motion \citep{Inoue11}.
Moreover, since in its the original formulation this model assumes the eddy motions to be isotropically distributed 
in the shell frame, another problem concerns the pulse shape, which is expected to be symmetric and thereby at odds with observations,
in which the decay is on average $\sim2$--$3$ times longer than the rise \citep{Norris96}.
This problem can be solved when reconnection takes place in ordered thin layers between
anti-parallel regions in the outflow (\citealt[hereafter BG15]{BeniaminiGranot15}). In this model, magnetic reconnection is
anisotropic in the shell comoving frame and isotropic in the emitter rest frame: the plasma in the reconnection region
is accelerated to $\gamma'$ along the reconnection layer in the shell comoving frame, so that the higher $\gamma'>1$,
the more anisotropic the emission in the shell frame.
They consider a thin shell as a reconnection layer that emits between $R_0$ and $R_0+\Delta R$.
The duration of the pulse is given by $T={\rm max}(\Delta T_r,\Delta T_\theta)$, where $\Delta T_r=\Delta R/2c\Gamma^2$
and $\Delta T_\theta=R_0/2c\Gamma^2\gamma'$ are the radial and the angular spreading times, respectively.
For $1/\gamma'\lesssim \Delta R/R_0$, higher values of $\gamma'$ yield narrower,
more symmetric, and more luminous (for a given total radiated energy) pulses. Hence, the higher $\gamma'$,
the stronger the weight of short-timescales on the PDS. 
On the spectral side, the radiation process is mainly synchrotron, and the expected synchrotron frequency, 
in the observer frame, is $\nu\propto\Gamma B'\gamma'^{2(1-\eta)/\eta}$ (where $B'$ is the magnetic field 
in the comoving shell frame). BG15 estimated $\eta\sim0.5$ to explain the narrowing of pulses with energy, 
$\Delta t\propto E^{-0.4}$ \citep{Fenimore95}, which implies $\nu\propto\Gamma B'\gamma'^2$. 
Assuming that $E_{\rm p,i}$ scales as the synchrotron frequency, it is $E_{\rm p,i}\propto\Gamma B'\gamma'^2$.
Our result can be interpreted as follows: soft GRBs with a steep PDS are dominated by the slow component
as the result of a relatively low relativistic magnetisation of the outflow at the dissipation site.
Consequently, emitters are accelerated to only mildly relativistic energies in the shell comoving frame.
By contrast, GRBs with shallower PDS mostly fit with 
{\sc pl} model are the result of more energetic magnetic reconnection events to $\gamma'\sim10$.
The large scatter of the $E_{\rm p,i}$--$\alpha$ correlation can be attributed to the distribution of bulk
Lorentz factor $\Gamma$, which varies from one GRB to another independently of $\gamma'$.
That the correlation is mainly driven by $\gamma'$ rather than $\Gamma$ can be explained by the quadratic
dependence on $\gamma'$ (provided that $B'$ is independent).
Moreover, a relatively narrow distribution in $\Gamma$ further helps to enhance
$\gamma'$ over $\Gamma$. This possibility seems plausible from the $\lesssim 0.2$~dex scatter of the $\Gamma$
distribution obtained from the afterglow peak interpreted as the fireball deceleration \citep{Ghirlanda12}.

A similar conclusion can be derived considering another model that invokes a magnetically dominated dissipation. 
In the ICMART model \citep{ICMART} magnetised shells ($\sigma>1$) collide at typical distances of classical
IS ($R\gtrsim 10^{15}$~cm), thus triggering cascades of turbulent reconnection.
Here each broad pulse corresponds to a single cascade of mini-emitters triggered by the collision of two
magnetised shells. The overall light curve of a multi-pulse GRB is explained by the occurrence of multiple independent
shocks after the fashion of classical IS. As such, the time history on timescales of several seconds is ruled by
the emission history of the inner engine.
Each reconnection event behaves like a mini-emitter with a given $\gamma'\propto \sqrt{1+\sigma}$ with respect
to the outflow frame \citep{Zhang14b}.
Interestingly, the calculated PDS of the individual light curves are fit with power laws, with occasional
breaks at high frequencies. The simulated PDS slopes span from $\alpha\sim1.1$ to
$\alpha\sim 2.1$. Spikier curves have shallower PDS, as expected since high frequencies carry relatively
more power than smoother light curves. Furthermore, the PDS slope is related in a simple way to
the Lorentz factor contrast $\gamma'/\Gamma$ for a fixed $\Gamma$: for low contrasts ($\sim0.01$) the PDS slope is around 2,
while  for $\gamma'/\Gamma\sim 0.1$ the PDS slope is around $1.1$--$1.2$. On average, a more magnetised outflow
(higher $\sigma$ means higher $\gamma'$) seems to show a stronger fast or spiky component. On the other hand, $E_{\rm p,i}$ 
depends on $\gamma'$ as well (assuming the same spectrum in the emitter comoving frame), and is therefore correlated
with the PDS slope.

\section{Conclusions}
\label{sec:con}
We studied for the first time the individual PDS of a sample of 123 long GRBs with known distance and well-measured time-average spectrum. Using Bayesian MCMC techniques, we modelled the individual PDS with either a
{\sc pl} or a {\sc bpl} model, depending on whether a break is required by a likelihood ratio test.
We found a highly significant correlation between time-average rest-frame peak energy $E_{\rm p,i}$ and
the PDS slope $\alpha$, so that shallower PDS are associated with high $E_{\rm p,i}$ values.
Moreover, GRBs with a break in the PDS, determined by the dominant timescale in the light curve,
tend to be relatively soft (i.e. low values of $E_{\rm p,i}$) and have steep PDS (above the frequency break).
We interpret this to mean that the variety of PDS is mainly driven by the relative strength of a fast component
in the light curves: the more variable the light curve at sub-second timescales, the shallower the PDS
and the less likely a break in the explored range $0.01$--$1$~Hz.

The only outlier GRB~060614 is noteworthy. This is known to be a peculiar nearby GRB with no associated 
SN that eludes the long-vs-short classification. Future data of bright short GRBs together with a more detailed
investigation of 060614 will help to clarify the behaviour of short GRBs in the $E_{\rm p,i}$--$\alpha$ plane and
add a new piece to the 060614 jigsaw.

We considered the main models that have been proposed in the literature to explain how or where GRB prompt emission
originates, with emphasis on the distance from the progenitor. Overall, the most natural way to explain the
$E_{\rm p,i}$--$\alpha$ correlation is invoking a common (kinematic) origin, such as Doppler boosting.
For photospheric models in which the dissipation takes place relatively close to the photosphere
($10^{12}$--$10^{13}$~cm), the fast component (when observed) keeps the memory of the variability imprinted
by the inner engine. One possible interpretation suggested by our results is related to the viewing 
angle, so that the more off axis the observer, the weaker the fast ($<1$~s) variability component in the light
curves, and the lower $E_{\rm p,i}$.

Moving outwards to distances in the range from $10^{14}$ up to $10^{16}$~cm, there are different mechanisms,
one of which is given by IS. Within the framework of the IS model, sub-second variability still reflects the
inner engine character through the distribution of bulk Lorentz factor $\Gamma$ of the wind of shells.
The presence of an $e^\pm$ photosphere acts like a low-pass filter for a wind of shells with relatively low
values of $\Gamma$ ($\approx$ a few tens), whereas for a wind of fast shells (e.g. $\Gamma$ of several hundreds)
this would not be the case. Unless a dependence on $\Gamma$ of microphysical parameters of
the shock physics is invoked or specific assumptions are made, such as continuous electron acceleration in the shock region,
the IS-predicted scaling $E_{\rm p,i}\propto\Gamma^{-2}$ seems hardly compatible with the
observed $E_{\rm p,i}$--$\alpha$ correlation.

Magnetic reconnection as the dissipation mechanism of the GRB prompt emission is another option.
The distance from the progenitor is comparably large as for the IS
model, but in this case, the fast component originates in situ and the source of energy is magnetic rather
than kinetic. The key to obtain sub-second variability at such large distance is the relativistic
($\gamma'\lesssim10$) motion of emitters within the comoving frame of relativistic ($\Gamma$) shells.
The details depend on the specific model: magnetic reconnection episodes could be triggered by shocks between
magnetised shells after the fashion of the baryonic shells of the IS model, as envisaged in the ICMART
model \citep{ICMART}, or according to an anisotropic emission within the comoving frame of the reconnection
layer \citep{BeniaminiGranot15}. In either case, the relative strength of the fast component in the observed
curves relates to the average Lorentz factor $\gamma'$ of the emitters within the comoving shell frame,
and hence to the magnetisation $\sigma$ itself of the shell, rather than its bulk Lorentz factor $\Gamma$.

In summary, the relative strength of the fast component, which positively correlates with $E_{\rm p,i}$
and determines a shallow PDS, can probe the magnetisation of the outflow.
In principle, this connection can be further tested by investigating other independent observables that
may be affected by high values of $\sigma$, such as a reverse shock in the early afterglow
\citep{ZK05,Japelj14}, or large-scale magnetic fields entrained in the ejecta as revealed
by prompt and early polarisation measurements
\citep{Mundell07sci,Steele09,Goetz09,Yonetoku12,Uehara12,Mundell13,Goetz14,Kopac15b}.
Ultimately, evidence for a highly magnetised jet can provide further support to scenarios where
the GRB progenitor is a newly born millisecond magnetar.

\begin{acknowledgements}
  We thank the anonymous referee for helpful comments that improved the paper.
  S.D., C.G., L.A., F.F. acknowledge support by PRIN MIUR project on ``Gamma Ray Bursts: from
  progenitors to physics of the prompt emission process'', P.~I. F. Frontera (Prot. 2009 ERC3HT).
\end{acknowledgements}

\bibliographystyle{aa}

\end{document}